\documentclass{elsart}

\usepackage{graphicx}
\usepackage{amsmath}
\usepackage{amssymb}

\begin{document}

\begin{frontmatter}

\title{The $^{8}$He and
$^{10}$He spectra studied in the $(t$,$p)$ reaction}

\author[dub]{M.S.~Golovkov,}
\author[dub,gsi,kur]{L.V.~Grigorenko,}
\author[dub]{G.M.~Ter-Akopian,\corauthref{cor1}}
\corauth[cor1]{corresponding author, e-mail: Gurgen.TerAkopian@jinr.ru}
\author[dub]{A.S.~Fomichev,}
\author[dub]{Yu.Ts.~Oganessian,}
\author[dub]{V.A.~Gorshkov,}
\author[dub]{S.A.~Krupko,}
\author[dub]{A.M.~Rodin,}
\author[dub]{S.I.~Sidorchuk,}
\author[dub]{R.S.~Slepnev,}
\author[dub]{S.V.~Stepantsov,}
\author[dub,hen]{R.~Wolski,}
\author[dub,chi]{D.Y.~Pang,}
\author[dub,cze]{V.~Chudoba,}
\author[kur]{A.A.~Korsheninnikov,}
\author[kur]{E.A.~Kuzmin,}
\author[kur,rik]{E.Yu.~Nikolskii,}
\author[kur]{B.G.~Novatskii,}
\author[kur]{D.N.~Stepanov,}
\author[gan]{P.~Roussel-Chomaz,}
\author[gan]{W.~Mittig,}
\author[bel]{A.~Ninane,}
\author[bru]{F.~Hanappe,}
\author[pas]{L.~Stuttg\'e,}
\author[rnfc]{A.A.~Yukhimchuk,}
\author[rnfc]{V.V.~Perevozchikov,}
\author[rnfc]{Yu.I.~Vinogradov,}
\author[rnfc]{S.K.~Grishechkin,}
\author[rnfc]{S.V.~Zlatoustovskiy}

\address[dub]{Flerov Laboratory of Nuclear Reactions, JINR, Dubna, 141980
Russia}
\address[gsi]{Gesellschaft f\"{u}r Schwerionenforschung mbH, Planckstrasse 1,
D-64291, Darmstadt, Germany}
\address[kur]{Russian Research Center ``The Kurchatov Institute'', Kurchatov sq.\ 1, 123182 Moscow, Russia}
\address[hen]{Henryk Niewidniczanski Institute of Nuclear Physics, Cracow,
Poland}
\address[chi]{Department of Technical Physics,
Peking University, Beijing 100871, People's Republic of China}
\address[cze]{Faculty of Nuclear Sciences and Physical Engineering, Czech Technical University, 115 19 Prague, Czech Republic}
\address[rik]{RIKEN, Hirosawa 2-1, Wako,Saitama 351-0198, Japan}
\address[gan]{GANIL, BP 5027, F-14076 Caen Cedex 5, France}
\address[bel]{Institut de Physique Nucl\'eaire and Centre de Recherches du Cyclotron, University of Louvain B1348 Louvain-la-Neuve, Belgium}
\address[bru]{Universit\'e Libre de Bruxelles, PNTPM, Bruxelles, Belgium}
\address[pas]{Institut de Recherches Subatomiques,IN2P3/Universit\'e Louis Pasteur, Strasbourg, France}
\address[rnfc]{RNFC -- All-Russian Research Institute of Experimental Physics, Sarov, Nizhni Novgorod Region, RU-607190 Russia}

\begin{abstract}
The low-lying spectra of $^8$He and $^{10}$He nuclei were studied in the
$^3$H($^6$He,$p$)$^8$He and $^3$H($^8$He,$p$)$^{10}$He transfer reactions. The
$0^+$ ground state (g.s.) of $^8$He and excited states, $2^+$ at $3.6-3.9$ MeV
and $(1^+)$ at $5.3-5.5$ MeV, were populated with cross sections of 200,
$100-250$, and $90-125$ $\mu$b/sr, respectively. Some evidence for $^8$He state
at about 7.5 MeV is obtained. We discuss a possible nature of the near-threshold
anomaly above 2.14 MeV in $^8$He and relate it to the population of a $1^-$
continuum (soft dipole excitation) with peak value at about 3 MeV. The lowest
energy group of events in the $^{10}$He spectrum was observed at $\sim 3$ MeV
with a cross section of $\sim 140$ $\mu$b/sr. We argue that this result is
possibly consistent with the previously reported observation of $^{10}$He, in
that case providing a new g.s.\ position for $^{10}$He at about 3 MeV.
\end{abstract}

\begin{keyword}
$^6$He, $^8$He beams, tritium gas target, resonance states, hyperspherical
harmonic method, soft dipole mode, neutron halo.

\PACS 25.10.+s -- Nuclear reactions involving few-nucleon systems, 24.50.+g --
Direct reactions, 25.55.Hp -- $^3$H, $^3$He, $^4$He induced reactions; transfer
reactions, 25.60.Ge -- Reactions induced by unstable nuclei; transfer reactions,
21.60.Gx -- Cluster models. \vspace{2mm}

\end{keyword}

\end{frontmatter}

\section{Introduction}

%###############################################################################

To study drip-line nuclei with large neutron excess one should either transfer
neutrons or remove protons or make multi-nucleon charge-exchange. Two-neutron
transfer from tritium provides here important opportunities connected with the
simplicity of reaction mechanism and simplicity of recoil particle (proton)
registration. This class of reactions remains practically not exploited in the
radioactive beam research. Availability of the unique cryogenic tritium target
\cite{yuk03} in the Flerov Laboratory of Nuclear Reactions (JINR, Dubna) makes
possible systematic studies of these reactions. The effectiveness of such an
approach in the investigation of exotic nuclei was demonstrated in the recent
studies of the $^5$H system \cite{gol04a,gol05b}.

Although $^{10}$He has been discovered more than a decade ago \cite{kor94}, very
limited information on this system is available. The ground state properties
were found in the $^2$H($^{11}$Li,$^{10}$He)$X$ reaction as
$E_{^{10}\mbox{\scriptsize He}}= 1.2(3)$, $\Gamma < 1.2$ MeV \cite{kor94}, and
in the $^{10}$Be($^{14}$C,$^{14}$O)$^{10}$He reaction as
$E_{^{10}\mbox{\scriptsize He}}= 1.07(7)$, $\Gamma = 0.3(2)$ MeV \cite{ost94}.
Here and below $E_{^{A}\mbox{\scriptsize He}}$ denotes the energy relative to
the lowest breakup threshold for the $A=\{6,8,10\}$ systems, while $E$ denotes
the excitation energy.

The $^{10}$He g.s.\ was theoretically predicted \cite{kor93} to be a narrow
three-body $^8$He+$n$+$n$ resonance with $E_{^{10}\mbox{\scriptsize He}} \sim
0.7-0.9$ MeV, $\Gamma \sim 0.1-0.3$ MeV and the valence neutrons populating
mainly the $[p_{1/2}]^2$ configuration. A widely discussed shell inversion
phenomenon in the $N=7$ nuclei became the source of new interest to $^{10}$He.
Possible existence of a virtual state in $^9$He was demonstrated in Ref.\
\cite{che01} and an \emph{upper} limit $a<-10$ fm was established for the
scattering length. Following this finding, the existence of a narrow
near-threshold $0^+$ state in $^{10}$He ($E_{^{10}\mbox{\scriptsize He}} =
0.05$, $\Gamma = 0.21$ MeV) with a structure $[s_{1/2}]^2$ was predicted in
Ref.\ \cite{aoy02} in addition to the $[p_{1/2}]^2$ $0^+$ state. It was
suggested in \cite{aoy02} that the ground state of $^{10}$He had not been
observed so far and the resonance at $\sim 1.2$ MeV is actually the first
excited state. The low-lying spectrum of $^9$He was revised in the recent
experiment \cite{gol07} resulting in a higher, than in the previous studies,
position of the $p_{1/2}$ state (experiment \cite{gol07} provided unique
spin-parity identification for the $^9$He states below 5 MeV). The presence of
the $s_{1/2}$ contribution is evident in the data \cite{gol07}, but the exact
nature of this contribution (virtual state or nonresonant $s$-wave continuum)
was not clarified and only a \emph{lower} limit $a>-20$ fm was set in this work.
This work triggered further theoretical research: problems with the
interpretation of the $^{10}$He spectrum and controversy between the $^9$He and
$^{10}$He data were demonstrated in Ref.\ \cite{gri08}.

This intriguing situation inspired us to revisit the $^{10}$He issue. The study
of the $^3$H($^8$He,$p$)$^{10}$He reaction was accompanied by the study of the
$^3$H($^6$He,$p$)$^8$He reaction providing a reference case of the relatively
well investigated $^8$He system.

%===============================================================================

\section{Experimental setup}

%===============================================================================

Experiments were performed using a 34 MeV/amu primary beam of $^{11}$B delivered
by the JINR U-400M cyclotron. The secondary beams of $^6$He and $^8$He nuclei
were produced by the separator ACCULINNA \cite{rod97} and focused in a 20 mm
spot on the target cell. For safety reasons, the main target cell, filled with
900 mPa tritium gas and cooled down to 28~K, was inserted into an evacuated
protective box. Thus, the target had twin entrance and exit windows sealed with
12.7 $\mu$m stainless steel foils. For 4 mm distance between the inner entrance
and exit windows the thickness of the tritium target was $2.0 \times 10^{20}$
cm$^{-2}$. Typical beam intensities incident on the target were $\sim
4\times10^4$ s$^{-1}$ for the $^6$He and $\sim 6\times10^3$ s$^{-1}$ for the
$^8$He projectile nuclei. The admixtures of other particles in the beams were no
more than $7 \%$ and the beam diagnostics completely eliminated them.

%-------------------------------------------------------------------------------
\begin{figure}[t]
\centerline{\includegraphics[width=0.48\textwidth]{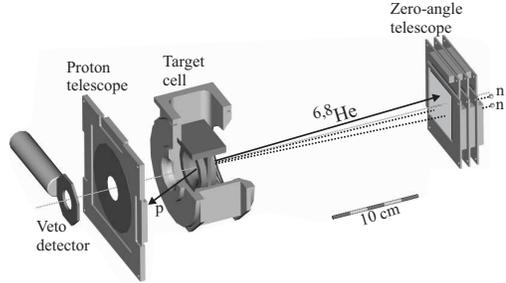}}
\caption{Experimental setup and kinematical diagram.}
\label{fig:setup}
\end{figure}
%-------------------------------------------------------------------------------

Experimental setup and kinematical diagram for the $^{3}$H($^6$He,$p$)$^8$He and
$^{3}$H($^8$He,$p$)$^{10}$He reactions are shown in Fig.\ \ref{fig:setup}. For
the small centre-of-mass system (cms) angles, where the maximal cross section is
expected, the protons fly in backward direction in the lab system. The residuals
($^{10}$He and $^8$He) and their decay products ($^8$He and $^6$He) are moving
in a relatively narrow angular cone in forward direction. Protons escaping back
from the target hit a telescope consisting of one 300 $\mu$m and one 1 mm thick
annular Si detectors. The active areas of these detectors had the outer and
inner diameters of 82 mm and 32 mm, respectively. The proton telescope was
installed 100 mm upstream of the target and covered an angular range of
$171^{\circ} - 159^{\circ}$ in lab system. The first detector was segmented in
16 rings on one side and 16 sectors on the other side and the second, 1 mm
detector was not segmented. A veto detector was installed upstream of the proton
telescope to alert to the signals generated by the beam halo.

Zero angle telescope for the $^{6}$He and $^{8}$He detection was installed on
the beam axis at a distance of 36.5 cm in the case of the $^6$He beam and at
28.8 cm in the experiment with the $^8$He beam. The telescope included six
squared ($60 \times 60$ mm) 1 mm thick detectors. The first two detectors of the
telescope were segmented in 16 strips each in vertical and horizontal
directions. All other detectors in the telescope were segmented in 4 strips in
the $^8$He run and in 16 strips in the $^6$He run.

A set of beam detectors was installed upstream of the veto detector (not shown
in Fig.\ \ref{fig:setup}). Two 0.5 mm plastic scintillators placed on a 8 m base
provided the particle identification and projectile energy measurement. The
overall time resolution was 0.5 ns. Beam tracking, giving a 1.5 mm resolution
for the target hit position, was made by two multiwire chambers installed 26 and
80 cm upstream of the target.

Particle identification in the proton telescope was not imperative because, due
to kinematical constraints, nothing but protons could be emitted in the backward
direction in these reactions. The main background source were protons
originating from the interactions of beam nuclei with the target windows. Test
irradiations done with empty target showed that this background was almost
completely eliminated when $p$-$^8$He or/and $p$-$^6$He coincidences were
considered. In the case of the $^{3}$H($^6$He,$p$)$^8$He reaction the detection
of the $p$-$^{8}$He coincidence events granted a selection for the reaction
channel populating the $^8$He g.s. For the decays of $^{10}$He and excited
$^{8}$He nuclei the respective $p$-$^{8}$He and $p$-$^{6}$He coincidence
information was used to clean the missing mass spectra and reconstruct the
charged fragment energy in the cms of $^{10}$He or $^8$He.

Array of 48 detector modules of the neutron time-of-flight spectrometer DEMON
\cite{til95} was installed in the forward direction at a distance of 3.1 m from
the target. In more rare events where triple $p$-$^{6}$He-$n$ coincidences were
detected the complete reaction kinematics was reconstructed.

For the $^{6}$He  and $^{8}$He  beams the projectile energies in the middle of
the tritium target were on average about 25 MeV/amu and 27.4 MeV/amu,
respectively; integral fluxes $2\times 10 ^{10}$ and $5\times 10 ^9$ were
collected. The missing mass spectra of $^{8}$He and $^{10}$He were measured up
to 14 MeV and 16 MeV, respectively. The upper limits were set by the low-energy
proton detection threshold. Monte Carlo (MC) simulations taking into account the
details of these experiments showed that a 450 keV (FWHM) resolution was
inherent to the $^8$He and $^{10}$He missing mass energy spectra obtained from
the data. The precision of the beam energy measurement made the most important
contribution to the error of the missing mass.

%===============================================================================

\section{$^3$H($^6$He,$p$)$^8$He reaction}

%===============================================================================
\begin{figure}[t]
\centerline{\includegraphics[width=0.43\textwidth]{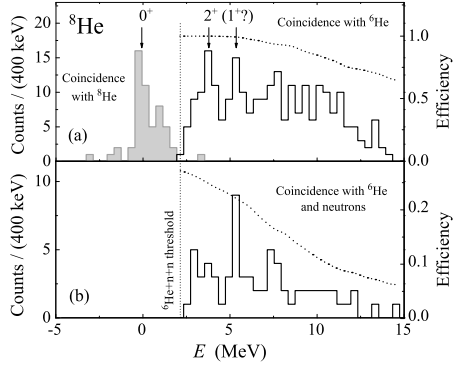}}
\caption{Missing mass spectrum of $^8$He. (a) The $p$-$^8$He and $p$-$^6$He
coincidence data were used to obtain the ground state peak and the excited
state spectrum, respectively. (b) Spectrum built for the $^8$He excited
states from the $p$-$^{6}$He-$n$ coincidence data. The efficiencies of the
$p$-$^{6}$He and $p$-$^{6}$He-$n$ coincidence registration are shown by dotted
curves (see the right axes in both panels).}
\label{fig:exp-8he}
\end{figure}
%-------------------------------------------------------------------------------

Missing mass spectra of $^8$He from the $^{3}$H($^6$He,$p$)$^8$He reaction are
presented in Fig.\ \ref{fig:exp-8he}. The peak corresponding to the $^{8}$He
g.s.\ is well seen in the $p$-$^{8}$He coincidence data. The tail visible in
Fig.\ \ref{fig:exp-8he} (a) on the right side of the g.s.\ peak was caused by
the pile-ups in the second (non-segmented) detector. Protons emitted from the
target with energy $\sim 8.5$ MeV correspond to the g.s.\ peak of $^8$He. They
passed through the 300 $\mu$m Si detector and were stopped in the second (1 mm)
detector of the proton telescope. The background signals arose here from the
beam halo particles [count rate of $(2-3)\times 10^3$ s$^{-1}$]. The veto
detector allowed taking away these events in the data analysis but the energy
resolution of the second detector was deteriorated. Operation conditions were
much better for the segmented 300 $\mu$m detector. The count rate per any of its
sectors was at least 10 times lower. Consequently, the background signals did
not cause the resolution deterioration when the $p$-$^{6}$He coincidences were
detected. In that case protons with energy $< 7.5$ MeV were emitted from the
target and practically all of them were stopped in the 300 $\mu$m detector.
Therefore, for the $^{8}$He excited states the stated 450 keV resolution is
valid.

There are two peaks apparent in the $^8$He excitation spectrum. We assign 2$^+$
to the $^8$He resonance at excitation energy $E\approx 3.6$ MeV. The 2$^+$
resonance with energy 3.57$\pm$0.12 MeV and width $\Gamma$=0.5$\pm$0.35 MeV was
for the first time unambiguously, and with that good precision, obtained in Ref.
\cite{kor93a}. Later on, this resonance was reported in a number of papers with
energies close to 3.6 MeV and widths $\Gamma \approx 0.5-0.8$ MeV (see, e.g.,
\cite[and Refs.\ therein]{til04,ska07}). We assume that the $E \approx 5.4$ MeV
peak seen in Fig.\ \ref{fig:exp-8he} is the $1^+$ resonance of $^8$He. The
ground for this assumption comes from various theoretical results (e.g.\
\cite{pie04,pan04,vol05}) stably predicting that in the $^8$He excitation
spectrum the next state after the $2^+$ should be the $1^+$ state. We note that
evidence for the peak at $E\sim 5-6$ MeV was found in Ref.\ \cite{kor93a}. The
$^8$He excited state at 5.4 MeV was recently reported also in Ref.\
\cite{ska07}. A rapid rise of the $^8$He spectrum at the $^6$He+$n$+$n$ decay
threshold is seen in Fig.\ \ref{fig:exp-8he}. This rise cannot be explained by
the left ``wing'' of the $2^+$ resonance. The peculiar threshold behaviour is
discussed in Section \ref{sec:threshold}. We note also that the spectra in Fig.\
\ref{fig:exp-8he} show some evidence for a $^8$He state at E$\approx$7.5 MeV.

In the $^{3}$H($^6$He,$p$)$^8$He reaction the population cross section for the
$^8$He g.s.,\ averaged in a range of $4^{\circ}-10^{\circ}$ of the reaction cms,
is found to be $\sim 200$ $\mu$b/sr. The observed threshold anomaly makes the
cross section derivation for the excited states of $^{8}$He more complicated
(and model dependent). The cross sections for the excited states are further
discussed in Sections \ref{sec:cross-sect} and \ref{sec:threshold}.

%===============================================================================

\section{$^3$H($^8$He,$p$)$^{10}$He reaction}

%===============================================================================

\begin{figure}[t]
\centerline{\includegraphics[width=0.48\textwidth]{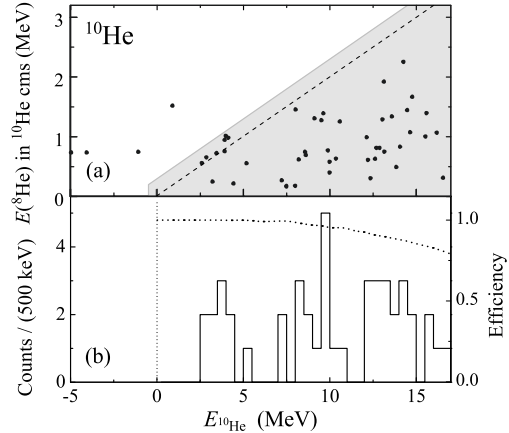}} \caption{(a)
Scatter plot showing the $^8$He energy observed in the $^{10}$He cms frame
versus the $^{10}$He missing mass energy. (b) Missing mass spectrum of
$^{10}$He. The $p$-$^{8}$He coincidence efficiency is shown by dotted curve (see
right axis).}
\label{fig:exp-10he}
\end{figure}

Data obtained for the $^3$H($^8$He,$p$)$^{10}$He reaction are shown in Fig.\
\ref{fig:exp-10he} (a) as a scatter plot $E(^8$He) vs.\
$E_{^{10}\mbox{\scriptsize He}}$, where $E(^8$He) is the energy of $^{8}$He in
the $^{10}$He cms. Condition $5E(^8$He$) \le E_{^{10}\mbox{\scriptsize He}}$
should be valid for the $^{10}$He decay. Therefore, $^{10}$He events should be
below the boundary shown by the dashed line in the scatter plot of Fig.\
\ref{fig:exp-10he} (a). The shaded area in Fig.\ \ref{fig:exp-10he} (a) extends
this boundary accounting for the experimental resolution. One can see that
practically all the events presented in Fig.\ \ref{fig:exp-10he} (a) fall into
the $^{10}$He locus indicating very clean background conditions. The missing
mass spectrum in Fig.\ \ref{fig:exp-10he} (b) was obtained projecting the events
confined in the $^{10}$He locus.

%-------------------------------------------------------------------------------
%-------------------------------------------------------------------------------

Not a single event was detected in the $^{10}$He spectrum below 2.5 MeV. This
imposes a stringent (one count corresponds to 14 $\mu$b/sr) limit on the
population cross section in the expected $^{10}$He ground state region at about
1.2 MeV \cite{kor94}. The lowest energy feature in the $^{10}$He spectrum is a
group of 10 events in between 2.5 and 5.5 MeV [see Fig.\ \ref{fig:exp-10he}].
This $\sim 3$ MeV group is well isolated from the rest of the spectrum and has a
typical resonant cross section ($\sim 140$ $\mu$b/sr averaged for cms angles
$3.5^{\circ}-9.5^{\circ}$), see estimates in Section \ref{sec:cross-sect}. Also,
this group has a distinct feature: the energy distribution of the $^{8}$He
fragments obtained in the $^{10}$He cms appears to be different from that in the
rest of events in the $^{10}$He spectrum. One can see in Fig.\
\ref{fig:exp-10he} (a) that within this group the $E(^8$He) energies are around
the maximal possible value. This means that the relative energy of the decay
neutrons for such events tends to zero. This could be evidence for some strong
specific momentum correlations or/and strong $n$-$n$ final state interaction in
this part of the $^{10}$He spectrum. We think that the $\sim 3$ MeV group of
events represents a resonant state for $^{10}$He; the possible nature of this
state is discussed in Section \ref{sec:interpr}.

%===============================================================================

\section{Cross section estimates}
\label{sec:cross-sect}

%===============================================================================

Both the $^8$He and $^{10}$He states were populated in our experiments by the
same ``dineutron'' transfer in the same kinematical conditions and, presumably,
by the same direct reaction mechanism. This fact makes it very probable that
spectroscopic information can be extracted from the cross sections in a
straightforward way. For theoretical estimates of the spectroscopic factors we
used somewhat extended phenomenological Cluster Oscillator Shell Model
Approximation (COSMA) of Ref.\ \cite{zhu94}. Within this model the g.s.\ wave
functions (WF) $\Psi^J$ of the $^{6,8,10}$He isotopes can be written as
\begin{eqnarray}
\Psi^0_{^6\mbox{\scriptsize He}} & = & \alpha_6 [p_{3/2}^2]_0 + \beta_6
[p_{1/2}^2]_0 \;, \nonumber  \\
\Psi^0_{^8\mbox{\scriptsize He}} & = & \alpha_8 [p_{3/2}^4]_0 + \beta_8
[p_{3/2}^2p_{1/2}^2] _0 \;, \nonumber \\
\Psi^{0(p)}_{^{10}\mbox{\scriptsize He}} & = & [p_{3/2}^4p_{1/2}^2] _0 \quad ,
\qquad \Psi^{0(s)}_{^{10}\mbox{\scriptsize He}}  =  [p_{3/2}^4s_{1/2}^2] _0 \;.
\label{eq:wfs-he}
\end{eqnarray}
The schematic notation $[l_{j}^n] _J$ denotes the Slater determinant of $n$
neutrons occupying $l_j$ orbital projected on the total spin $J$ and normalized.
The $\alpha$-particle is considered to be an inert core and it is omitted in the
notation. In the original paper \cite{zhu94} only the $\alpha_8$ configuration
in Eq.\ (\ref{eq:wfs-he}) was considered.

The model looks very schematic. However, it lists \emph{all} the possible
$p$-shell configurations, representing the dominant part of the WF.
Particularly, for the $^6$He g.s.\ coefficients $\alpha_6$, $\beta_6$ can be
inferred from the three-cluster model calculations \cite{dan91}
\[
\alpha_6 = 0.926 \,, \quad \alpha^2_6 = 0.86  \,, \quad
\beta_6 = 0.226 \,, \quad \beta^2_6 = 0.05\;,
\]
exhausting $91\%$ of the WF normalization (the corresponding $79\%$ of
$K=2,\,L=0$ and $12\%$ of $K=2,\,L=1$ components are considered). The simplified
$^{6}$He WF can also be used with only $p_{3/2}$ configuration ($\alpha_6 = 1$,
$\beta_6 = 0$) to test the sensitivity to the $^6$He structure. Assuming the
$^8$He WF (\ref{eq:wfs-he}) is normalized, we end up with only one unknown
parameter $\beta_8$ in the model.

The cluster overlaps for the $^{8,10}$He WFs within this model are:
\begin{eqnarray}
\langle \Psi^0_{^8\mbox{\scriptsize He}} | \Psi^0_{^6\mbox{\scriptsize He}}
\rangle & = &  \frac{\alpha_6 \beta_8} {\sqrt{6}} \, [p_{1/2}^2]_0 +
\frac{\beta_6 \beta_8 - \alpha_6 \sqrt{1-\beta_8^2}} {\sqrt{6}} \, [p_{3/2}^2]_0
\, , \nonumber \\
\langle \Psi^{0(p)}_{^{10}\mbox{\scriptsize He}} | \Psi^0_{^8\mbox{\scriptsize
He}} \rangle & = & \frac{\sqrt{1-\beta_8^2}} {\sqrt{15}} \, [p_{1/2}^2]_0 -
\frac{\beta_8} {\sqrt{15}} \, [p_{3/2}^2]_0 \,, \nonumber \\
\langle \Psi^{0(s)}_{^{10}\mbox{\scriptsize He}} | \Psi^0_{^8\mbox{\scriptsize
He}} \rangle & = & \frac{\sqrt{1-\beta_8^2}} {\sqrt{15}} \, [s_{1/2}^2]_0  \,.
\label{eq:overlap}
\end{eqnarray}
Using spin algebra and Talmi coefficients, the overlaps of the shell model
configurations with the ``dineutron'' $nn$ being in the $s$-wave motion relative
to the core are obtained as
\[
\langle [p^2_{3/2}]_0 | nn \rangle = \sqrt{\frac{2}{6}} \;, \quad \langle
[p^2_{1/2}]_0 | nn \rangle = \sqrt{\frac{1}{6}} \;,\quad \langle [s^2_{1/2}]_0 |
nn \rangle = 1 \;.
\]
Dineutron here is the the two neutrons with angular momentum and total spin
equal to zero represented by minimal oscillator. The spectroscopic weight of the
$^{6}$He g.s.\ configuration in the $^{8}$He WF is obtained by Eq.\
(\ref{eq:overlap}) as
\begin{equation}
\Bigl \| \langle \Psi^0_{^8\mbox{\scriptsize He}} | \Psi^0_{^6\mbox{\scriptsize
He}} \rangle \Bigr \|  =   \frac{1}{6} \biggl[ \alpha^2_6 \beta^2_8 + \Bigl(
\beta_6 \beta_8 - \alpha_6 \sqrt{1-\beta_8^2}\Bigr) ^2 \biggr] \, .
\label{eq:norm-6he}
\end{equation}
For the reactions studied in this work a reasonable estimate of the cross
section ratio $\sigma_{10}/\sigma_8$ for the $^{10}$He  and $^8$He g.s.\
population is the ratio of the dineutron spectroscopic factors. They are found
as
\begin{eqnarray}
S_8^{nn} &= &\frac{4!}{2!2!} \,\langle \Psi^0_{^8\mbox{\scriptsize He}} |
\Psi^0_{^6\mbox{\scriptsize He}}, nn \rangle
^2 \nonumber \\
 & = & \frac{1}{6} \Bigl[\alpha_6 \Bigl(\beta_8-\sqrt{2(1-\beta_8^2)}\Bigr) +
\sqrt{2} \beta_6 \beta_8 \Bigr]^2 \,, \nonumber \\
S_{10}^{nn}(p) &= & \frac{6!}{2!4!} \, \langle
\Psi^{0(p)}_{^{10}\mbox{\scriptsize He}} | \Psi^0_{^8\mbox{\scriptsize He}}, nn
\rangle ^2 = \frac{1}{6} \Bigl[\sqrt{1-\beta_8^2} - \sqrt{2}  \beta_8 \Bigr]^2
\,,\nonumber \\
S_{10}^{nn}(s) & = & \frac{6!}{2!4!} \, \langle
\Psi^{0(s)}_{^{10}\mbox{\scriptsize He}} | \Psi^0_{^8\mbox{\scriptsize He}}, nn
\rangle ^2 = 1-\beta_8^2 \,.\nonumber
\label{eq:spec-fac}
\end{eqnarray}
%

%===============================================================================
%
\begin{figure}[t]
\centerline{ \includegraphics[width=0.43\textwidth]{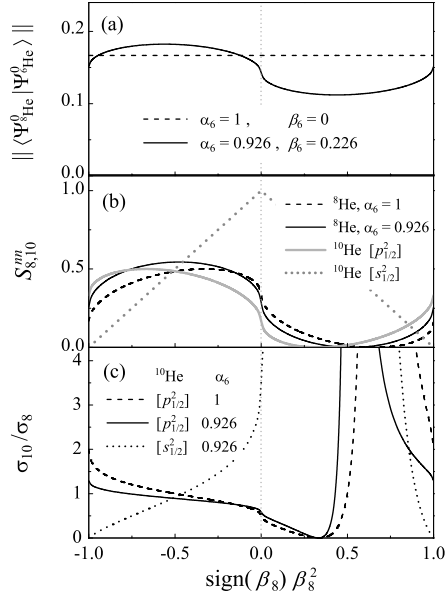}  }
\smallskip
\caption{(a) Spectroscopic weight (\ref{eq:norm-6he}) of the $^6$He g.s.\
configuration in $^8$He WF for the simplified ($\alpha_6=1$) and realistic
structures of $^6$He. (b) Two-neutron spectroscopic factors in $^8$He and
$^{10}$He. (c) Estimated cross section ratio for the $^{10}$He and $^8$He g.s.\
population in the $(t,p)$ reaction.}
\label{fig:spec-fac}
\end{figure}
%
%===============================================================================

The spectroscopic information obtained in the model is illustrated by Fig.\
\ref{fig:spec-fac}. In the region $\beta_8 >0$ the cross section ratio is
changing dramatically [Fig.\ \ref{fig:spec-fac} (c)]. However, this region is
presumably unphysical. In this region the weight of the dineutron configuration
in $^{8,10}$He is minimal [Fig.\ \ref{fig:spec-fac} (b)] and the weight of the
$^6$He g.s.\ configuration in $^8$He is minimal as well [Fig.\
\ref{fig:spec-fac} (a)]. These configurations are expected to be maximized by
the variational procedure as they are energetically highly preferable. Simple
heuristic considerations show that the $\beta_8$ coefficient should be confined
by condition $\beta_8<0$ [to maximize attractive $(ls)$ interaction] and
$-0.5<\,$sign$(\beta_8)\beta^2_8<-0.3$ [to maximize pairing].

\begin{enumerate}

\item For a reasonable weight of coefficient $\beta_8$ (for example,
$-0.5<\,$sign$(\beta_8)\beta^2_8<0$) the
population of the $[s^2_{1/2}]$ state in $^{10}$He is expected to be larger than
the $[p^2_{1/2}]$ state.

\item Population cross section for the $^{10}$He $[p^2_{1/2}]$ state can not
differ strongly from that obtained for the $^8$He g.s. For the realistic
structure of $^6$He the values lying in a range of $\sigma_{10}/\sigma_8 \sim
0.6-1.3$ are expected.

\item Population rate for the 3 MeV group of events in $^{10}$He
is found consistent with the resonant cross section estimated for the population
of the $p$-wave state.  Coefficient $\beta_8$ can be obtained from the
experimentally measured cross sections for the population of $^{8}$He and
$^{10}$He ground states: $\beta^2_8 \approx 0.1^{+0.3}_{-0.1}$. In this work
such a derivation is methodologically clean as both cross sections are obtained
in the same experimental conditions.

\item Note that the model proposed here (with neutrons situated only in the
$p$-shell) shows that the basic dynamics of the $^8$He system strongly limits
the possible range of the $^6$He g.s.\ configuration weight in the $^8$He
structure [see Fig.\ \ref{fig:spec-fac} (a)]. This implies that the weights
corresponding to the $^6$He g.s.\ and $^6$He$(2^+)$ configurations in the
structure of $^8$He have only a weak dependence on the $[p^4_{3/2}]_0$ and
$[p^2_{3/2}p^2_{1/2}]_0$ configuration mixing.

\item The spectroscopic factor for processes with the disintegration of
$^8$He in $^6$He(g.s)+2n continuum is connected with the weight in Eq.\
(\ref{eq:norm-6he}) by relation
\[
S^{2n}_8 = 6 \; \Bigl \| \langle \Psi^0_{^8\mbox{\scriptsize He}} |
\Psi^0_{^6\mbox{\scriptsize He}} \rangle \Bigr \| \;.
\]
A discrepancy can be seen in Ref.\ \cite{chu05} between the experimentally
obtained $S^{2n}_8=1.3(1)$ and theoretical ``shell model'' value given as 1/6
(see Table 1 in\ \cite{chu05}). The values obtained in our model vary between
0.8 and 1.1 (depending on the $\beta_8$ value) in a good agreement with the
experiment of Ref.\ \cite{chu05}.

\end{enumerate}

%===============================================================================

\section{Possible nature of the threshold state in $^8$He}
\label{sec:threshold}

%===============================================================================

In the missing mass spectrum of $^8$He (see Fig.\ \ref{fig:exp-8he}) attention
is attracted by a steep rise ensuing straight from the three-body $^6$He+$n$+$n$
threshold. The lowest known resonant state of $^8$He is $2^+$ at $E=3.57$ MeV
\cite{kor93a}, $\Gamma=0.5-0.7$ MeV. It decays sequentially via the $^7$He
ground state resonance ($3/2^-$ at $E_{^{7} \mbox{\scriptsize He}}=0.445$ MeV,
$\Gamma=0.15$ MeV) by a $p$-wave neutron emission. This guarantees negligible
population of the continuum below $\sim 0.6$ MeV where decay takes place in a
``three-body regime'', $\sigma \sim E_{^{8} \mbox{\scriptsize He}}^4$. Above
that energy, population probability transfers to the ``two-body $p$-wave
regime'', $\sigma \sim E_{^{8}\mbox{\scriptsize He}}^{3/2}$. Consequently, the
low-energy tail of the $2^+$ state can not be responsible for the near threshold
events.

The only plausible source of the low-energy events, we have found, is the
population of the E1 (means $1^-$) continuum. Theoretical studies of such
continuum populated in reactions \cite{dan98,ers01,ers04} show that the profile
of the $1^-$ cross section typically well resembles the profile of the
electromagnetic strength function $dB_{E1}/dE$. Such functions for spatially
extended halo systems could provide very low-energy peak --- the so called soft
dipole mode
--- even without the formation of any $1^-$ resonant state.

We estimate the E1 strength function for the $^8$He$\rightarrow
^6$He+$2n$ dissociation using the model developed in \cite{gri06}.
For the WF with outgoing asymptotic
\begin{equation}
\Psi_E^{(+)}(\mathbf{X},\mathbf{Y}) = \int d\mathbf{X}' d \mathbf{Y}' \,
G_E^{(+)}(\mathbf{XX}',\mathbf{YY}') \, \hat{D} \, \Psi_{\mbox{\scriptsize
g.s.}}(\mathbf{X},\mathbf{Y}) \,,
\label{eq:psi-plus}
\end{equation}
generated by the dipole operator $\hat{D}$, acting on the g.s.\ WF
$\Psi_{\mbox{\scriptsize g.s.}}$, the E1 strength function is found as
\[
\frac{dB_{E1}}{dE} = \frac{2J_f+1}{2J_i+1} \, \frac{X^2}{2\pi} \, \mbox{Im}
\left[\int d\Omega_x \int d\mathbf{Y} \,\Psi_E^{(+)\dagger} \,
\frac{\nabla_x}{M_x} \, \Psi_E^{(+)} \right] \biggr |_{X \rightarrow \infty} .
\]
Vectors $\mathbf{X}$ and $\mathbf{Y}$ are Jacobi coordinates for the
$^{6}$He-$n$ and ($^{6}$He-$n$)-$n$ subsystems, respectively. Estimating the
dipole strength for the light $p$-shell nuclei we can well take into account
only the $[p^2]\rightarrow [sp]$ transitions and neglect the $nn$ interactions
and $s$-wave interaction between the core and neutron (unless the latter is not
strongly attractive). In this approximation the three-body Green's function (GF)
has a simple analytical form
\[
G_E^{(+)}(\mathbf{XX}',\mathbf{YY}') = \frac{1}{2\pi i} \! \int \! \!
dE_{^{7}\mbox{\tiny He}}  G_{E_{^{7}\mbox{\tiny He}}}^{(+)}  \!
(\mathbf{X},\mathbf{X}')G_{E-E_{^{7}\mbox{\tiny He}}}^{(+)} \!
(\mathbf{Y},\mathbf{Y}'),
\]
where $G_{E-E_{^{7}\mbox{\scriptsize He}}}^{(+)}$ is a free motion GF in the $Y$
subsystem, and the GF in the $X$ subsystem corresponds to the $p$-wave
continuum with the $^7$He g.s. $3/2^-$ resonance at $E_{^{7}\mbox{\scriptsize
He}}=0.445$ MeV.

%===============================================================================
%
\begin{figure}[t]
\centerline{ \includegraphics[width=0.40\textwidth]{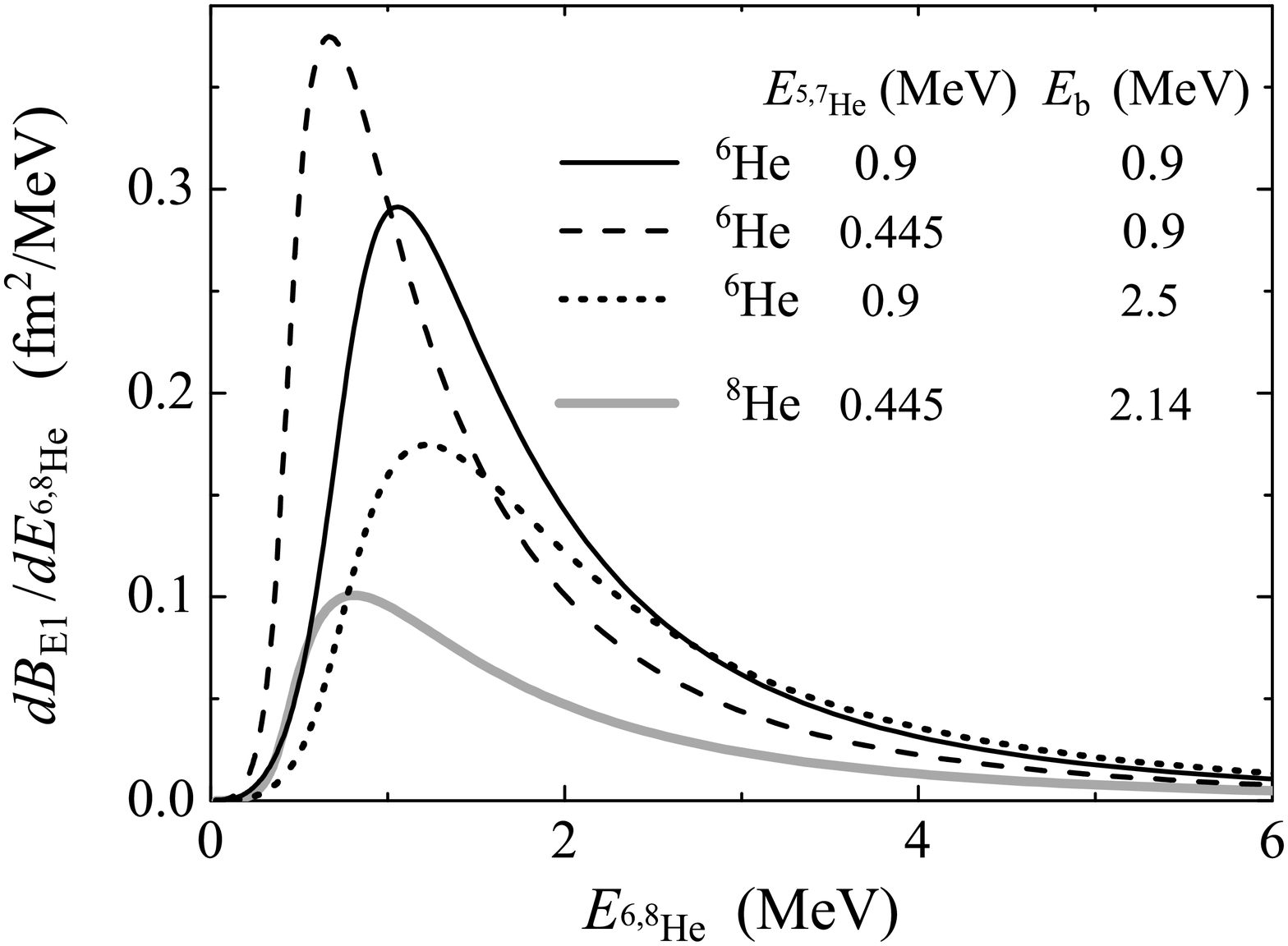}  }
\smallskip
\caption{E1 strength function for $^{6}$He and $^{8}$He. The dashed and dotted
curves show calculations done for $^{6}$He with unrealistic parameters
demonstrating the trends in the strength functions behaviour with the parameter
variation.}
\label{fig:dbde}
\end{figure}
%
%===============================================================================

The results of the model calculations, including the $^{6}$He test, are shown in
Fig.\ \ref{fig:dbde}. The estimated $^{6}$He strength function giving peak at
about 1.1 MeV is in a reasonable agreement with the complete three-body
calculations \cite{dan98} giving peak at about 1.3 MeV. It can be seen that the
strength function profile in $^{6}$He is sensitive to two main aspects of the
dynamics. (i) Energy of the resonance ground state in the $p$-wave subsystem:
dashed curve shows that the strength function peak is shifting to the
\emph{lower} energy if the $^{5}$He $3/2^-$ state is artificially shifted from
the experimental $E_{^{5}\mbox{\scriptsize He}}=0.9$  MeV position to the
\emph{lower} 0.445 MeV. (ii) ``Size'' of the ground state WF: dotted curve shows
the strength function peak shifting to \emph{higher} energy if we artificially
overbound the $^{6}$He g.s.\ WF to $E_b=2.5$ MeV instead of 0.9 MeV
\emph{decreasing} its radial extent. When we turn from $^{6}$He to $^{8}$He
these dynamical trends work in the opposite directions and largely compensate
each other (the $^{8}$He g.s.\ is more ``compact'' than the $^{6}$He g.s., but
the $^{7}$He g.s.\ resonance is lower than the $^{5}$He g.s.\ resonance). As a
result we find the strength function peak position in $^{8}$He to be somewhat
lower than respective position in $^{6}$He. This indicates that in $^{8}$He,
where the $2^+$ state is significantly higher than in $^{6}$He, the
lowest-energy feature in the continuum could be the $1^-$ excitation.

%===============================================================================
%
\begin{figure}[t]
\centerline{ \includegraphics[width=0.44\textwidth]{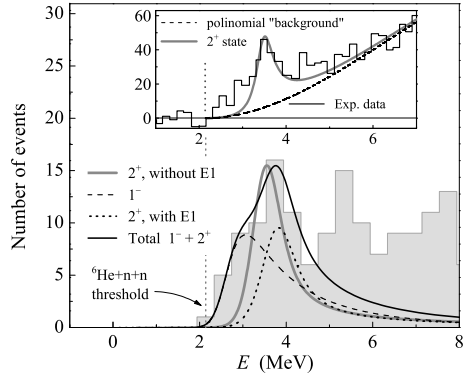}  }
\smallskip
\caption{Experimental spectrum of Fig.\ \ref{fig:exp-8he} is compared to the
theoretical profile of the $2^+$ state with and without a possible contribution
of the $1^-$ continuum. Theoretical curves are convoluted with the experimental
resolution. The experimental data of \cite{boh95} are shown in the inset.
Dashed curve is the result of polynomial ``background'' subtraction and solid
gray line is the same as the solid gray line in the main panel (convoluted with
the experimental resolution of \cite{boh95}).}
\label{fig:dbde-comp}
\end{figure}
%
%===============================================================================

The behaviour of the cross section with the estimated E1 component taken into
account is shown in Fig.\ \ref{fig:dbde-comp}. The $2^+$ state profile is given
here by the standard R-matrix expression for the $p$-wave decay via the $^{7}$He
g.s.\ providing the widths $\Gamma=0.56-0.82$ MeV for excitation energies $E=
3.6-3.9$ MeV (the reduced width is taken as Wigner limit). Without E1
contribution the data are in agreement with the standard position ($E\approx
3.6$ MeV) of the $2^+$ state, but the near threshold behaviour of the cross
section can not be reproduced. The $2^+$ population cross section in this case
can be estimated as $\sim 250$ $\mu$b/sr. The addition of the $1^-$ contribution
allows to reproduce the low-energy part of the spectrum much better. In that
case we can allow up to $60\%$ feeding to the $1^-$ continuum. Then we get $\sim
100$ $\mu$b/sr for the $2^+$ population and have to shift to about $E\approx
3.9$ MeV the position of this state.

The proposed significant contribution of the $1^-$ cross section is not
absolutely unexpected and never seen phenomenon. For example, the experimental
spectrum from paper \cite{boh95} is shown in the inset to Fig.\
\ref{fig:dbde-comp}. Inspected around the $^{6}$He+$n$+$n$ threshold ``on a
large scale'' it shows the same presence of the low-energy intensity which can
not be attributed to the tail of the $2^+$ state. Strong population of the E1
continuum in $^8$He by nuclear processes has been demonstrated in a comparison
made for the nuclear and Coulomb dissociation data \cite{mar01,mei02}. However,
in the interpretation of the data presented in \cite{mar01,mei02} the idea was
accepted that the E1 cross section in $^{8}$He should peak at \emph{higher}
energy than in $^{6}$He (maximum at about $E_{^{8}\mbox{\scriptsize He}} \approx
2$ MeV above the threshold). This idea is based on the argument (ii) discussed
above (smaller size of $^8$He compared to $^6$He); actual situation appears to
be more complicated. As a result the authors of \cite{mar01,mei02} have had to
position the $2^+$ state \emph{below} the E1 peak. Consequently, they had to
ascribe to it a very low excitation energy 2.9 MeV (compared to about 3.6 MeV in
the other recent works). The assumption of the very low-energy soft E1 peak in
$^{8}$He would probably allow to explain in a more natural way the data from
\cite{mar01,mei02}. Also, there exists a large uncertainty in the definition of
the ``standard'' position of the $2^+$ state in $^{8}$He ($2.7-3.6$ MeV, see
Ref.\ \cite{til04}). We think that a significant component of the disagreement
among different experimental works could be connected with the possibility that
the $2^+$ state is typically observed in a mixture with the $1^-$ contribution.
Correlation measurements could clarify this situation.

%===============================================================================

\section{Interpretation of the $^{10}$He spectrum}
\label{sec:interpr}
%===============================================================================

There is an evident discrepancy between the group of events at about 3 MeV
observed in this experiment and the recognized position of $^{10}$He g.s.\ at
about 1.2 MeV. A possible explanation is that an excited state of $^{10}$He was
observed in our work and the ground state was not populated for some reason. We,
however, find a different explanation preferable.

%===============================================================================
%
\begin{figure}[t]
\centerline{ \includegraphics[width=0.45\textwidth]{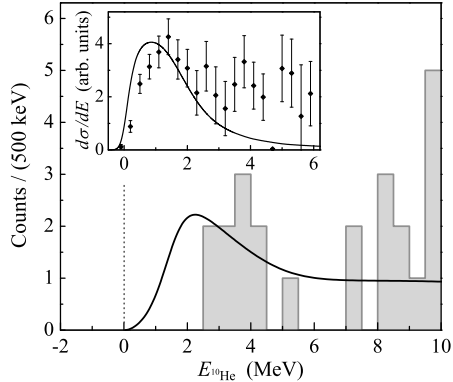}  }
\smallskip
\caption{Missing mass spectrum of $^{10}$He compared to the predictions of
Ref.\ \cite{gri08} for the $^{10}$He g.s.\ with the $[p^2_{1/2}]$ structure.
The inset shows the experimental data of Ref. \cite{kor94} compared to the
theoretical spectrum obtained with the same Hamiltonian as in the main panel
but for the different reaction mechanism.}
\smallskip
\label{fig:th-comp}
\end{figure}
%
%===============================================================================

There are two important problems, pointed by theoreticians, in the
interpretation of the $^{10}$He spectrum. (i) Possible existence of a
near-threshold $0^+$ state with the $[s^2_{1/2}]$ structure, due to the shell
inversion phenomenon \cite{aoy02}. In this case we would have two $0^+$ states
in the low-energy continuum of $^{10}$He, nearby each other. The $[s^2_{1/2}]$
state is predicted in \cite{gri08} to have very specific properties (tentatively
assigned as ``three-body virtual state'') and it distorts strongly the
higher-lying spectrum associated with the $[p^2_{1/2}]$ $0^+$ state. At first
blush it is not impossible that the $[s^2_{1/2}]$ $0^+$ state is not populated
in our experiment. (ii) Reaction mechanism issue was pointed in Ref.\
\cite{gri08}. The most clear observation of the $^{10}$He g.s.\ was made so far
in the experiment with the $^{11}$Li beam \cite{kor94}. It was shown in
\cite{gri08} that, in contrast to the typical transfer reactions, the
experiments with the $^{11}$Li beam can provide very specific signal for the
$[p^2_{1/2}]$ $0^+$ state: in the $^{11}$Li case the spectrum is shifted
downwards due to the abnormal size of the halo component of the $^{11}$Li WF.

Let us consider the second issue first. The measured missing mass spectrum of
$^{10}$He is shown in Fig.\ \ref{fig:th-comp} in comparison with the
experimental data \cite{kor94} and calculations \cite{gri08} taking into
account the reaction mechanisms in both cases. It is clear that the
calculations are somewhat overbound ($\sim 0.5-0.7$ MeV), but otherwise
consistent with the data in both cases. It has also been shown in Sec.\
\ref{sec:cross-sect} that the absolute cross section value for the 3 MeV group
of events is quantitatively consistent with the population of a $p$-wave state.
We can conclude here that it is very probable that the 1.2 MeV peak observed in
Ref.\ \cite{kor94} and the 3 MeV peak in our work represent the same state. It
should be emphasized that the calculated peak energy for the $(t$,$p)$ reaction
cross section is consistent with the resonance properties inferred from the
$S$-matrix in \cite{gri08}: the eigenphase for $3 \rightarrow 3$ scattering is
passing $\pi/2$ at about the peak energy. Therefore, the peak energy observed
in the transfer reaction could provide a better access to the $^{10}$He
properties.

%===============================================================================
%
\begin{figure}[t]
\centerline{ \includegraphics[width=0.40\textwidth]{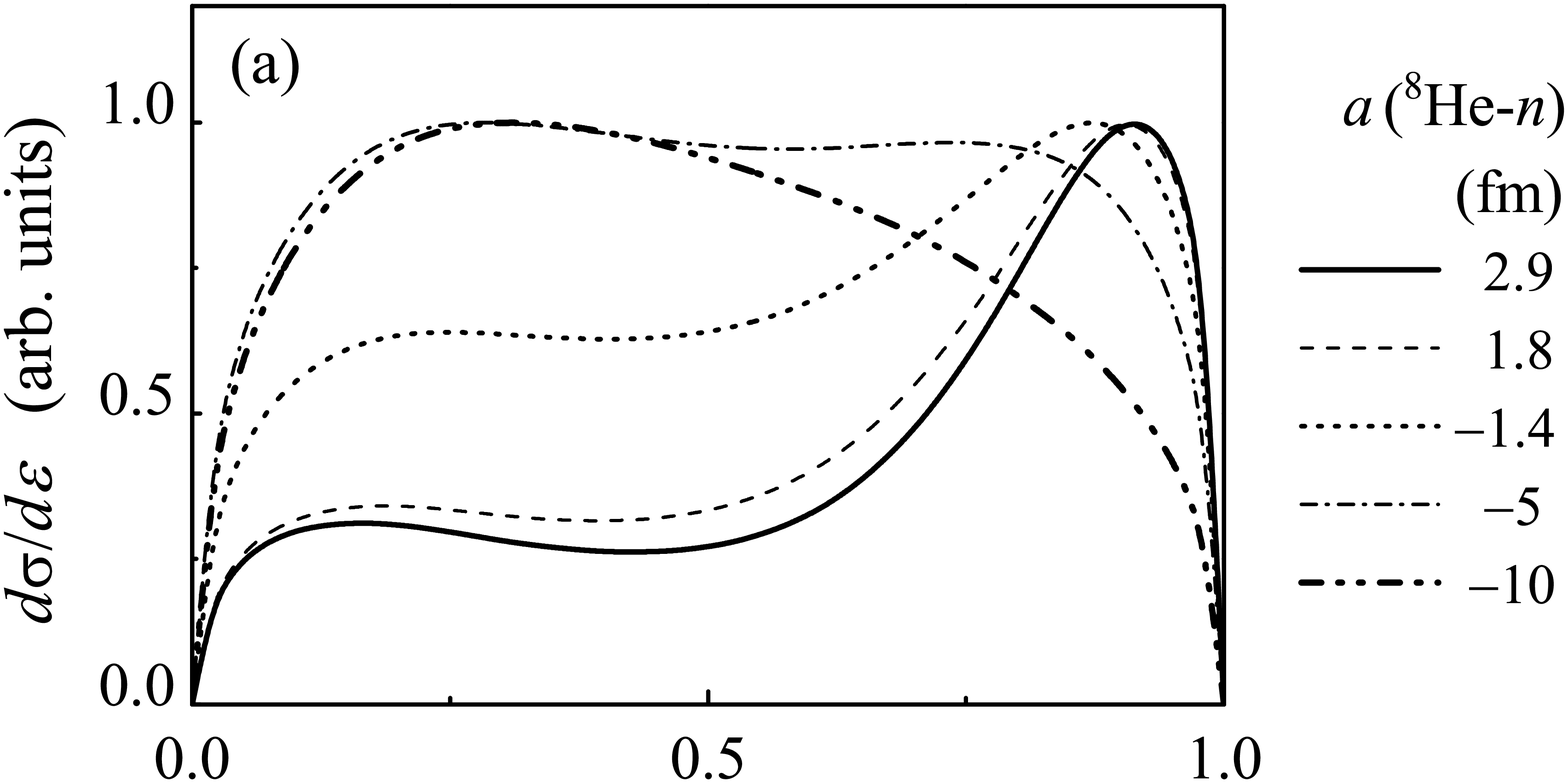}  }
\centerline{ \includegraphics[width=0.40\textwidth]{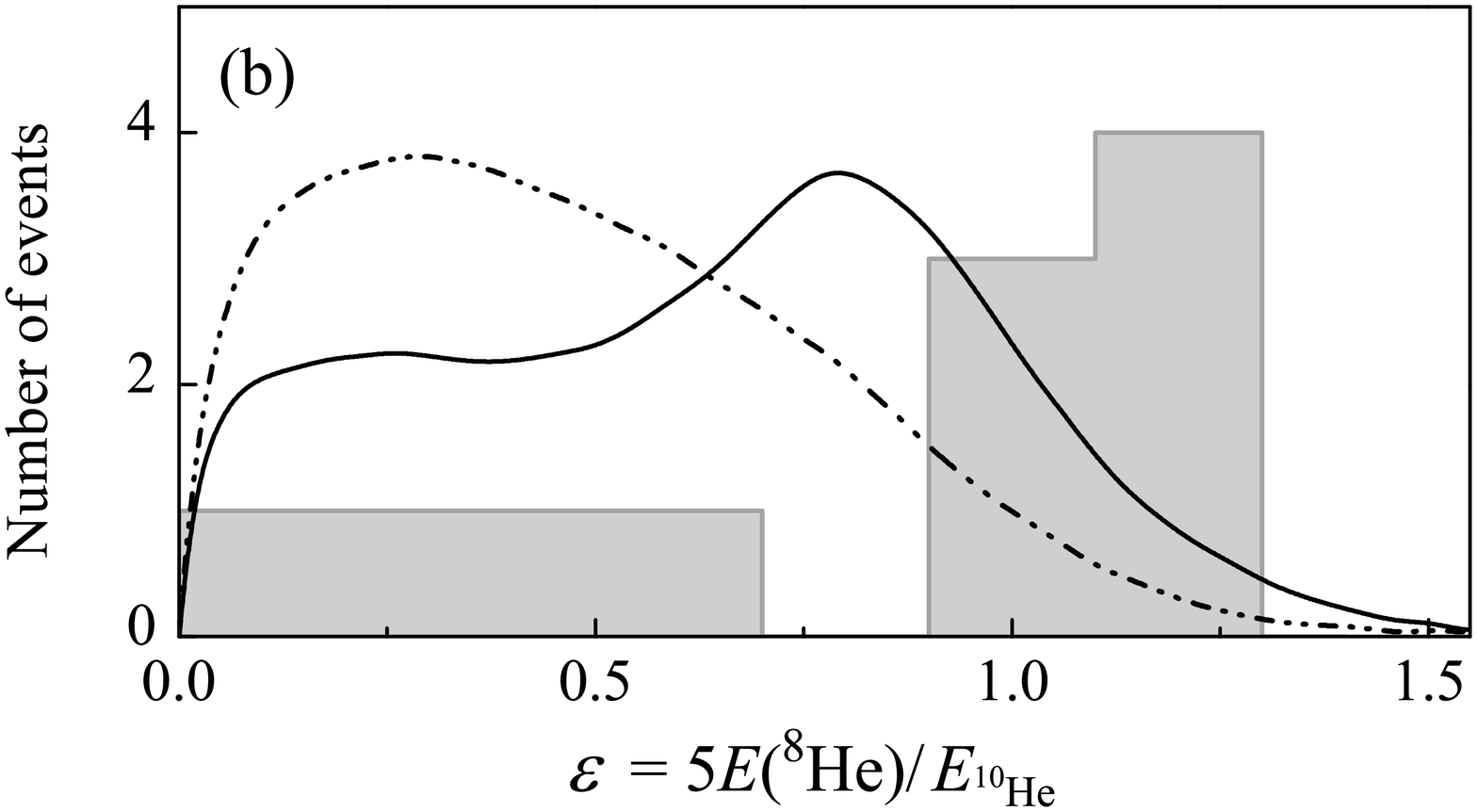}  }
\smallskip
\caption{Energy distribution of $^8$He in the $^{10}$He cms frame. (a)
Calculated in Ref.\ \cite{gri08} for different interactions in the $^8$He-$n$
channel. Corresponding scattering lengths are shown in the legend; positive $a$
values correspond to repulsive interactions. Solid curve
corresponds to the theoretical missing mass spectrum given in Fig.\
\ref{fig:th-comp} ($^{10}$He g.s.\ with $[p^2_{1/2}]$ structure). (b) $E(^8$He)
distribution observed in the present work for the 3 MeV group of events is
shown as a gray histogram. Theoretical curves are the same as in panel (a)
convoluted by the MC procedure with the experimental resolution.}
\label{fig:th-en-dis}
\end{figure}
%
%===============================================================================

Now we return to the first issue. Is it possible that the theoretically
predicted in \cite{aoy02} low-lying $0^+$ state with the $[s^2_{1/2}]$ structure
exists, but it is not populated in our reaction? It was shown in \cite{gri08}
that the expected $0^+$ states with the $[s^2_{1/2}]$ and $[p^2_{1/2}]$
structures would interfere strongly. The momentum distributions for the
$[p^2_{1/2}]$ state were predicted to be strongly different in the cases when
there \emph{is} a  $[s^2_{1/2}]$ state below it and when there \emph{is no} such
state. This point is illustrated in Fig.\ \ref{fig:th-en-dis} (a) for different
interactions in the $^{8}$He-$n$ $s$-wave channel (the positive values of
scattering length indicated for two curves in Fig.\ \ref{fig:th-en-dis} (a)
imply that repulsive interaction takes place in the $s$-wave state). In Ref.\
\cite{gri08} the cases of $a<-5$ fm in $^9$He correspond to the formation of
extremely sharp near threshold $0^+$ $^{10}$He states. Otherwise, there is only
the $[p^2_{1/2}]$ state at $\sim 2.4$ in the $^{10}$He continuum. It can be seen
in Fig.\ \ref{fig:th-en-dis} (b) that only scattering lengths $a \ge -5$ fm (and
hence no $[s^2_{1/2}]$ state) are qualitatively consistent with our data. Thus
the data favour the situation of the $[p^2_{1/2}]$ ground state of $^{10}$He. In
this way our data also indirectly lead to contradiction with the $^{8}$He-$n$
scattering length limit $a<-10$ fm claimed in Ref.\ \cite{che01}.

The interpretation proposed above is very nonorthodox and is based, at the
moment, on the limited statistics data. However, alternatively we face a problem
to explain why the ``real'' ground state was not observed in our experiment
despite the very low cross section limit achieved ($\sigma_{10} < 14$ $\mu$b/sr)
and the estimates of Section \ref{sec:cross-sect} indicating large population
probability for possible $[s^2_{1/2}]$ state.

%===============================================================================

\section{Conclusion.}

%===============================================================================

In this work we studied the  $^8$He and $^{10}$He spectra in the same $(t,p)$
transfer reaction. This allowed us, when interpreting the data, to be free in
our speculations of the reaction mechanism peculiarities. We think that our
results are not in contradiction with the previous works done on these nuclei in
the sense of the data, however, making various theoretical estimates we arrived
at different conclusions on several issues.

\begin{enumerate}

\item The ground $0^+$ and the excited $2^+$, $(1^+)$ states of $^8$He are
populated with cross sections 200, $\sim 100-250$, and $\sim 90-125$ $\mu$b/sr.
The presence of near-threshold events at about $E\sim 2.14$ MeV can be an
evidence for the formation of the soft dipole mode in the $^8$He continuum. The
generation of such a mode with the very low peak energy ($E_{^8\mbox{\scriptsize
He}}\sim 0.9$ MeV, $E\sim 3$ MeV) in nuclear reactions could possibly be an
explanation to the respective controversial features of the other $^8$He data as
well.

\item The population cross section of the 3 MeV peak in $^{10}$He
$\sigma_{10}=140(30)$ $\mu$b/sr is consistent with the estimated resonance cross
section for the population of the $^{10}$He state with the $[p^2_{1/2}]$
structure. The weight $\beta_8 \approx 0.1^{+0.3}_{-0.1}$ of the
$[p^2_{3/2}p^2_{1/2}]$ configuration in $^8$He was inferred from the
$\sigma_{10}/\sigma_8$ ratio.

\item According to the calculations of Ref.\ \cite{gri08} the 3 MeV peak
position obtained here for the $^{10}$He g.s.\ is in agreement with the 1.2 MeV
position found in Ref.\ \cite{kor94}, if one takes into account the peculiar
reaction mechanism for the $^{11}$Li beam used in \cite{kor94}. If this
interpretation is valid, a new ground state energy of about 3 MeV should be
established for $^{10}$He since the peak position obtained in the transfer
reaction corresponds to the $S$-matrix pole position, while for reactions with
$^{11}$Li there is a strong difference.

\item The absence of the near-threshold state in $^{10}$He, predicted to have a
$[s^2_{1/2}]$ structure \cite{aoy02} imposes, according to calculations
\cite{gri08}, a stringent limit $a>-5$ fm on the $^8$He-$n$ scattering length.
This is in contradiction with the existence of a virtual state in $^9$He,
declared to have $a<-10$ fm in Ref.\ \cite{che01}.

\end{enumerate}

Further measurements of a similar style are desirable. This would allow to
reveal the potential of correlation measurements for such complicated systems
and to resolve the interesting problems outlined in this work.

%===============================================================================
%
\section{Acknowledgments.}
%
%===============================================================================
%  05-02-16404  5H-9H by bombardment of H, D targets by 6He, 8He
%  05-02-17535  Danilin

We are grateful to Profs.\ B.V.\ Danilin, S.N.\ Ershov, and M.V.\ Zhukov for
illuminating discussions. The authors acknowledge the financial support from the
INTAS Grants No.\ 03-51-4496 and No.\ 05-100000-8272, Russian RFBR Grants Nos.\
05-02-16404, 08-02-00089 and 05-02-17535 and Russian Ministry of Industry and
Science grant NS-1885.2003.2. Support provided for this work by the Department
of Science and Technology of South Africa is acknowledged.

%###############################################################################

%###############################################################################

\end{document}